\documentclass[conference]{IEEEtran}
\usepackage{graphicx}
\usepackage{amsmath,bbm,amssymb,amsfonts,amstext,amsopn}
\usepackage{cite}
\usepackage{balance}
\usepackage{url}
\usepackage{epsfig}
\usepackage{setspace}
\usepackage{stmaryrd}
\usepackage{psfrag}	
\usepackage{multirow}
\usepackage{float}
\usepackage[process=auto]{pstool}
\usepackage{etoolbox}
\usepackage{algorithm}
\usepackage{algorithmic}
\allowdisplaybreaks

\newtheorem{prop}{Proposition}
\newtheorem{corol}{Corollary}

\newtoggle{OneColumn}

\toggletrue{OneColumn}

\iftoggle{OneColumn}{%
  
}{%
}

\EndPreamble
\begin{document}

\title{Diffusive Molecular Communications \\   with Reactive Signaling}

\author{
Vahid Jamali\dag, Nariman Farsad\ddag, Robert Schober\dag, and Andrea Goldsmith\ddag  \\
\dag University of Erlangen-Nuremberg, Germany
\quad
\ddag Stanford University,  USA\vspace{-0.3cm}
}

\maketitle

\begin{abstract}
This paper focuses on molecular communication (MC) systems  where the signaling molecules may participate in a reversible bimolecular reaction in the channel. The motivation for studying these MC systems is that they can realize the concept of constructive and destructive signal superposition, which leads to  favorable properties such as inter-symbol interference (ISI) reduction and avoiding environmental contamination due to continuous release of molecules into the channel.  This work first derives the maximum likelihood (ML) detector for a binary MC system with reactive signaling molecules under the assumption that the detector has perfect knowledge of the ISI. The performance of this genie-aided ML detector yields an upper bound on the performance of any practical detector. In addition, two suboptimal detectors of different complexity are proposed. The proposed ML detector as well as one of the suboptimal detectors require the channel response (CR) of the considered MC system. Moreover, the CR is needed for the performance evaluation of all proposed detectors.  However, analyzing  MC with reactive signaling is challenging since the underlying partial differential equations that describe the reaction-diffusion mechanism are coupled and non-linear. Therefore,  an algorithm is developed in this paper for efficient computation of the CR to any arbitrary transmit symbol sequence. The accuracy of this algorithm is validated via particle-based simulation.   Simulation results using the developed CR algorithm show that the performance of the proposed suboptimal detectors can approach that of the genie-aided ML detector. Moreover, these results show that MC systems  with reactive signaling have superior performance relative to those with non-reactive signaling due to the reduction of ISI enabled by the chemical reactions.
\end{abstract}


\section{Introduction}
Recent advances in biology, nanotechnology, and medicine have given rise to the need for communication in nano/micrometer scale dimensions \cite{Nariman_Survey}. In nature,  a common strategy for communication between nano/microscale entities such as bacteria, cells, and organelles (i.e., components of
cells) is diffusive molecular communication (MC) \cite{CellBio}. In contrast to conventional wireless communication systems that encode data into electromagnetic
waves, MC systems embed data in the characteristics of  signaling molecules such as their type and concentration. Therefore,
diffusive MC has been considered as a bio-inspired approach for communication between small-scale
nodes for applications where conventional wireless communication may be inefficient or infeasible.

One characteristic of MC is that the receiver always observes a \textit{constructive} superposition of the number of molecules  released in previous symbol intervals or by different transmitters since the numbers of molecules cannot be negative. This feature leads to  several undesirable effects. First, many concepts in conventional communications that rely on both constructive and destructive superposition of signals, such as precoding, beamforming, and orthogonal sequences, are not  applicable in MC. Second, the release of signaling molecules in consecutive symbol intervals introduces significant inter-symbol interference (ISI) as the channel impulse response of MC channels is heavy-tailed. Third, if molecules are continuously released, particularly into a bounded environment, the concentration of the signaling molecules increases over time and contaminates the environment. 

One solution to cope with these challenges is to use enzymes to degrade the signaling molecules in the environment \cite{Adam_Enzyme}. It has been shown in \cite{Adam_Enzyme} that ISI is significantly reduced if enzymes are uniformly present in the environment. However, having uniformly distributed enzymes in the environment has two main drawbacks. First,   degradation of the signaling molecules via enzymes cannot be controlled, which may hurt performance.  Second, the ISI reduction comes at the expense of  reducing the peak concentration of the signaling molecules observed at the receiver.  In \cite{Nariman_Acid}, the authors proposed to employ acids and bases for  signaling. This MC system has the advantage that the release of the molecules can be controlled by the transmitter and acids and bases can react to cancel each other out. Note that the reaction of an acid and a base produces water and salt, and hence the contamination of the environment by signaling molecules is avoided. Moreover, the use of acids and bases implies the destructive and constructive superposition of signaling molecules \textit{in the channel} (not at the receiver) which can be exploited to reduce ISI. In fact, the effectiveness of this reactive signaling for ISI reduction  was \textit{experimentally} verified in \cite{Nariman_AcidBasePlatform}. These advantages of the MC system in \cite{Nariman_Acid,Nariman_AcidBasePlatform} motivate us to consider MC systems with reactive signaling molecules in this paper.

The main challenge in analyzing MC with reactive signaling is that the underlying partial differential equations (PDEs) that describe the reaction-diffusion mechanism are \textit{coupled} and \textit{non-linear}. A closed-form solution to these equations has not yet been found, which had led to various approximations  \cite{NonlinearPDE_Debnath}. For instance,  in \cite{Adam_Enzyme}, the spatial and temporal distribution of the enzyme concentration was assumed to be constant to obtain an approximate solution. However, for MC systems in which the transmitter releases reactive signaling molecules into the channel, the concentrations of the molecules are temporally and spatially non-uniform and hence the constant distribution assumption does not hold. In the absence of closed-form solutions, numerical methods are commonly used to solve reaction-diffusion equations   in the chemistry and physics literature   \cite{PDE_numerical}.  This approach was applied to MCs in \cite{Nariman_Acid} where the authors employed a finite difference method (FDM) to solve the reaction-diffusion equation for a one-dimensional environment.  Another approach to compute the \textit{expected} concentrations of molecules is to average many realizations of concentrations obtained via a stochastic reaction-diffusion simulation  \cite{ReactionDiffSim,Chou_MasterEq,Adam_AcCoRD}. However, the computational complexity of these numerical and simulation methods is very high.  In \cite{Reza_Reaction}, data is encoded in the concentration difference of two types of molecules and it is shown that assuming identical diffusion coefficients for both types of signaling molecules, the resulting PDE for the concentration difference is linear. However, the statistical model for the difference of the observed molecules is still a function of the concentrations of both types of molecules.

In this paper, we consider a binary MC system that employs two types of molecules for signaling where the signaling molecules may participate in a \textit{reversible bimolecular reaction}, such as the acid and base reaction in \cite{Nariman_Acid}. The considered reversible bimolecular reaction involves two reactions with different rates, namely the reaction of two reactant molecules that yields a product molecule and the decomposition of this product molecule into the two reactant molecules. 
 Moreover, we assume an unbounded environment and a passive receiver for simplicity. For this system, we first derive a genie-aided maximum likelihood (ML) detector which assumes perfect knowledge of previous symbols. We also propose two suboptimal detectors with different complexities. The proposed ML detector and one of the suboptimal detectors require computation of the channel response (CR) of the considered MC system. Moreover, the CR is needed for the performance evaluation of all proposed detectors. 
The CR is complicated by the non-linearity that arises due to the bimolecular reaction, hence it must be characterized for all possible sequences of molecules released by the transmitter into the channel. To address the complexity of this characterization, we develop an  algorithm for efficient computation of the CR of the considered MC system. This algorithm is faster than the numerical methods that discretize both space and time since it efficiently exploits the simplifying characteristics of an unbounded environment and a passive receiver,  and computes the concentrations of the molecules analytically in each time step.  The accuracy of the proposed algorithm for CR computation is validated using particle-based simulation. Moreover, we show that unlike the MC system in \cite{Adam_Enzyme},  ISI is reduced in the considered MC system without reducing the peak of the CR. Finally, simulations using the proposed CR algorithm show superior bit error rate (BER) performance of MC systems with reactive signaling compared to those with non-reactive signaling due to reduced~ISI. 


\section{System Model}\label{Sec:SysMod}

\begin{figure}
  \centering
 \scalebox{0.95}{
\pstool[width=1\linewidth]{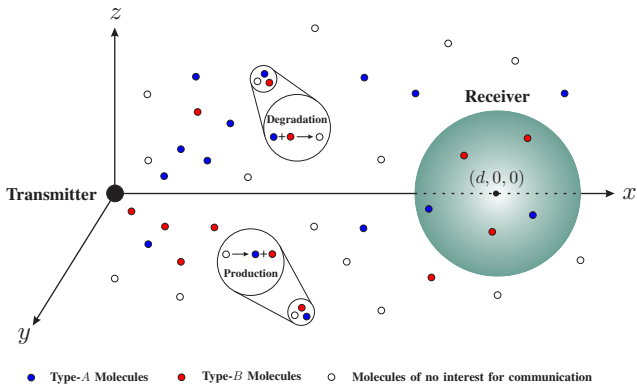}{
\psfrag{T}[l][c][0.7]{\textbf{Transmitter}}
\psfrag{R}[c][c][0.7]{\textbf{Receiver}}
\psfrag{D}[c][c][0.7]{$(d,0,0)$}
\psfrag{X}[c][c][1]{$x$}
\psfrag{Y}[c][c][1]{$y$}
\psfrag{Z}[c][c][1]{$z$}
\psfrag{P}[c][c][0.45]{\textbf{Degradation}}
\psfrag{G}[c][c][0.45]{\textbf{Production}}
\psfrag{F1}[l][c][0.5]{\textbf{Type-$A$ Molecules}}
\psfrag{F2}[l][c][0.5]{\textbf{Type-$B$ Molecules}}
\psfrag{F3}[l][c][0.5]{\textbf{Molecules of no interest for  communication}}
} }  \vspace{-0.3cm}
\caption{Schematic illustration of the considered MC system with reactive signaling.\vspace{-0.3cm} }
\label{Fig:SysMod}
\end{figure}

The considered MC system consists of a transmitter, a receiver, and a channel which are introduced in detail in the following, see Fig.~\ref{Fig:SysMod}. 

\subsection{Transmitter}\label{Sec:Transmitter}

We assume a point-source transmitter located at the origin of the Cartesian coordinate system, i.e., $(0,0,0)$.  The transmitter employs two types of molecules for signaling, namely type-$A$ and type-$B$ molecules. In particular, the transmitter releases $N_i^{\mathrm{tx}}$ type-$i$ molecules into the channel at time instances $t\in\mathcal{T}_i,\,i\in\{A,B\}$. By properly defining $\mathcal{T}_i$, different modulation schemes can be accommodated, e.g., molecule shift-keying (MoSK) and pulse position modulation (PPM). In this paper, we focus on the following modulation scheme. Let $s[k]\in\{0,1\}$ denote the binary symbol at the $k$-th symbol interval. For binary zero, $s[k]=0$, the transmitter releases $N_A^{\mathrm{tx}}$ type-$A$ molecules at the beginning of the symbol interval and $N_B^{\mathrm{tx}}$ type-$B$ molecules at time $\tau_0$ seconds after the start of the symbol interval. In a similar manner, for binary one, $s[k]=1$, the transmitter releases $N_B^{\mathrm{tx}}$ type-$B$ molecules at the beginning of the symbol interval and $N_A^{\mathrm{tx}}$ type-$A$ molecules at time $\tau_1$ seconds after the start of the symbol interval. In particular, we choose $\tau_0$ ($\tau_1$) as the peak of the CR assuming instantaneous release of \textit{only} $N_A^{\mathrm{tx}}$ ($N_B^{\mathrm{tx}}$) type-$A$ (type-$B$) molecules at $t=0$. For this modulation scheme, we obtain 
\begin{IEEEeqnarray}{lll} \label{Eq:Ti}
\mathcal{T}_A=\left\{t|t=(k-1)T^{\mathrm{symb}}+s[k]\tau_1,\,\,\forall k\right\} \IEEEyesnumber\IEEEyessubnumber \\
\mathcal{T}_B=\left\{t|t=(k-1)T^{\mathrm{symb}}+(1-s[k])\tau_0,\,\,\forall k\right\}, \IEEEyessubnumber
\end{IEEEeqnarray}
where $T^{\mathrm{symb}}$ denotes the length of a symbol interval.  The advantage of the above modulation scheme and the  choice of $\tau_i$ in reactive MC systems is that, unlike the MC system in \cite{Adam_Enzyme}, where the reduction of ISI comes at the expense of reducing the peak of the CR, here, the transmitter releases the second type of molecule only when the receiver has already seen the expected peak concentration of the first type of molecule.  Hence, the second release does not have an impact on the peak concentration of the first release.

\subsection{Channel}\label{Sec:Channel}

We assume an unbounded three-dimensional environment. 
The type-$A$ and type-$B$ molecules released by the transmitter diffuse in the environment with diffusion coefficients $D_A$ and $D_B$, respectively, and may participate in the following biomolecular reaction
\begin{IEEEeqnarray}{lll} \label{Eq:Reaction}
A+B \underset{k_b}{\overset{k_f}{\rightleftharpoons}} \varnothing,
\end{IEEEeqnarray}
where $k_f$ and $k_b$ denote the forward reaction rate in molecule$^{-1}$m$^{3}$s$^{-1}$ and backward reaction rate in s$^{-1}$, respectively. Moreover, symbol $\varnothing$ denotes chemical species which are of no interest for communication. Note that (\ref{Eq:Reaction}) includes the reactions considered in \cite{Nariman_Acid,Reza_Reaction}. Moreover, if  type-$B$ molecules  represent enzymes and only type-$A$ molecules are used for signaling, (\ref{Eq:Reaction}) includes the degradation reaction in \cite{Adam_Enzyme} when the enzyme concentration is constant everywhere and is much larger than the concentration of the type-$A$ molecules such that the reaction in (\ref{Eq:Reaction}) does not change the enzyme concentration.

\subsection{Receiver}\label{Sec:Receiver}

For simplicity, we assume a passive receiver at distance $d$ centered at  point $\mathbf{d}=(d,0,0)$ which is able to count the number of type-$A$ and type-$B$ molecules within its volume. Let $\bar{y}_A(t)$ and $\bar{y}_B(t)$ denote the  \textit{expected} numbers of type-$A$ and type-$B$ molecules observed at the receiver at time $t$, respectively, due to release of a known sequence of numbers of molecules by the transmitter. We refer to $\bar{y}_i(t),\,\,i\in\{A,B\}$, as the CR of the considered MC system. We assume perfect synchronization between the transmitter and the receiver \cite{ICC2017_MC_IEEE}. Note that depending on the length of the symbol interval, the receiver may observe molecules released by the transmitter in multiple previous symbol intervals, i.e., ISI may exist.  Let $L$ be the length of the channel memory\footnote{Theoretically, the memory length of the considered MC channel is infinite; however, from a practical point-of-view, the effect of the previous symbols becomes negligible after several symbol intervals.} and $\mathbf{s}\in\{0,1\}^{L-1}$ denote the vector of $L-1$ previously transmitted symbols which is referred to as the ISI-causing sequence\footnote{For notational simplicity, we drop the symbol index $k$ in the remainder of the paper.}. Therefore, given $\mathbf{s}$ and $s$, the number of  type-$i$ molecules counted at the receiver at sample time $t_s$ is modelled as
\begin{IEEEeqnarray}{lll} \label{Eq:Poisson_ISI}
y_i \sim \mathcal{P}\left(s\bar{y}_i^{(1)}(\mathbf{s})+(1-s)\bar{y}_i^{(0)}(\mathbf{s})\right),\quad i\in\{A,B\},
\end{IEEEeqnarray}
where $\mathcal{P}(\lambda)$ denotes a Poisson random variable (RV) with mean $\lambda$. Moreover, $\bar{y}_i^{(s)}(\mathbf{s})$ is $\bar{y}_i(t_s)$ under the condition that the symbol in the current symbol interval is $s\in\{0,1\}$ and the ISI-causing sequence is $\mathbf{s}\in\{0,1\}^{L-1}$. We note that the Poisson model in (\ref{Eq:Poisson_ISI}) is an approximation which has been shown to be accurate for the reaction-diffusion processes in the chemistry and physics literature \cite{PoissonGardiner,CoxNatureCommun}. In Section~\ref{Sec:SimResult}, we will validate the Poisson model   in (\ref{Eq:Poisson_ISI})  using the particle-based simulator developed in Appendix~\ref{App:SimParticle}.  

Note that due to the reaction process, the CR of the considered MC system $\bar{y}_i(t)$ is a non-linear function of the transmitted data symbols. Therefore, we cannot simply compute the CR for one shot transmission and use convolution to capture the effect of the ISI \cite{TCOM_MC_CSI}. 
In particular, to fully characterize the average behavior of the system, one has to compute the CR for both symbol hypotheses $s\in\{0,1\}$ and all $2^{L-1}$ possible ISI-causing sequences. In Section~\ref{Sec:Analysis}, we derive the optimal genie-aided ML detector and two suboptimal detectors for this MC system. Note that the CR is needed for both the ML and one of the suboptimal detectors, and is also required for  performance evaluation of all the proposed detectors. Therefore, in Section~\ref{Sec:Model}, we derive an efficient numerical algorithm for computation of the CR for any arbitrary sequence of transmitted symbols.

\section{Detection Methods for Binary Modulation}\label{Sec:Analysis}

In this section, we derive the genie-aided ML detector for binary modulation assuming the ISI is known. This provides an upper bound on performance for any practical detector. Subsequently, we propose two suboptimal  practical detectors of different complexity.

\subsection{Optimal Genie-Aided ML Detector}\label{Sec:MLDetector}

In the following, we focus on symbol-by-symbol detection. We consider a genie-aided ML detector that assumes  perfect knowledge of the ISI-causing sequence is available at each symbol interval.  In particular, the genie-aided ML detection problem for the considered transmission scheme is given by
\begin{IEEEeqnarray}{lll} \label{Eq:MLprob}
\hat{s}^{\mathrm{ml}} &= \underset{s\in\{0,1\}}{\mathrm{argmax}}\,\,\Pr(y_A,y_B|s,\mathbf{s}) \nonumber \\
&\overset{(a)}{=} \underset{s\in\{0,1\}}{\mathrm{argmax}}\,\, f_{\mathcal{P}}(y_A|s,\mathbf{s})f_{\mathcal{P}}(y_B|s,\mathbf{s}),
\end{IEEEeqnarray}
where $\Pr(\cdot)$ denotes probability and $f_{\mathcal{P}}(x)=\frac{\lambda^x e^{-\lambda}}{x!}$ is the probability mass function (PMF) of a Poisson RV with mean~$\lambda$. Equality $(a)$ follows from the fact that conditioned on $\bar{y}_i^{(s)}(\mathbf{s})$ and $(s,\mathbf{s})$, RVs $y_A$ and $y_B$ are independent. 
The optimal detector is given in the following proposition.

\begin{prop}\label{Prop:MLDetector}
The genie-aided ML detector as a solution of (\ref{Eq:MLprob}) is given by
\begin{IEEEeqnarray}{lll} \label{Eq:MLdetector}
\hat{s}^{\mathrm{ml}} = \begin{cases}
0, \quad &\mathrm{if}\,\, y_A \geq \alpha(\mathbf{s}) y_B + \beta(\mathbf{s}) \\
1, \quad &\mathrm{otherwise},
\end{cases}
\end{IEEEeqnarray}
where $\alpha(\mathbf{s})=\frac{1}{\gamma(\mathbf{s})}\log\Big(\frac{\bar{y}_B^{(1)}(\mathbf{s})}{\bar{y}_B^{(0)}(\mathbf{s})}\Big)$, 
$\beta(\mathbf{s})=\frac{1}{\gamma(\mathbf{s})}\big(\bar{y}_A^{(0)}(\mathbf{s})+\bar{y}_B^{(0)}(\mathbf{s})-\bar{y}_A^{(1)}(\mathbf{s})-\bar{y}_B^{(1)}(\mathbf{s})\big)$, and 
$\gamma(\mathbf{s}) = \log\Big(\frac{\bar{y}_A^{(0)}(\mathbf{s})}{\bar{y}_A^{(1)}(\mathbf{s})}\Big)$.
\end{prop}
\begin{IEEEproof}
The proof is given in Appendix~\ref{App:Prop_MLDetector}.
\end{IEEEproof}

Note that the detector in Proposition~\ref{Prop:MLDetector} requires the CR  $\bar{y}_i^{(s)}(\mathbf{s})$ for every sequence $(s,\mathbf{s})$. Therefore, in Section~\ref{Sec:Model}, we derive an algorithm for efficient  computation of the CR.

\subsection{Suboptimal Detectors}\label{Sec:Suboptimal}
In the following, we propose two suboptimal detectors.

\subsubsection{Suboptimal Detector~1} The genie-aided detector in Proposition~\ref{Prop:MLDetector} assumes perfect knowledge of the ISI-causing sequence which is not available in practice. However, the receiver can employ the detector in Proposition~\ref{Prop:MLDetector} and use its estimates of the previous symbols as the ISI-causing sequence.  This leads to a suboptimal detector which we refer to as ``ML detector with estimated ISI". We show in Section~\ref{Sec:SimResult} that the performance of the ML detector with estimated ISI is very close to the performance upper bound provided by the genie-aided ML detector.

\subsubsection{Suboptimal Detector~2} Recall that the main motivation for introducing the adopted modulation scheme was to reduce ISI. Assuming that ISI is sufficiently suppressed and that $D_A=D_B$ and $N_A=N_B$ hold, we propose the following simple detector
\begin{IEEEeqnarray}{lll} \label{Eq:MLdetector_simple}
\hat{s}= \begin{cases}
0, \quad &\mathrm{if}\,\, y_A \geq  y_B \\
1, \quad &\mathrm{otherwise}.
\end{cases}
\end{IEEEeqnarray}
 The suboptimal detector in (\ref{Eq:MLdetector_simple}) does not need knowledge of the CR which makes it suitable for MC systems with limited computational capabilities.


\section{Channel Response Computation  \\ for MC Systems with Reactive Signaling}\label{Sec:Model}
In this section, we first formally present the problem statement for CR computation. Next, we derive a numerical algorithm for computing the CR and discuss its complexity with respect to the available methods for CR computation.

\subsection{Problem Statement}\label{Sec:Problem}
Let $C_A(\mathbf{r},t)$ and  $C_B(\mathbf{r},t)$ denote the concentration of type-$A$ and type-$B$ molecules at point $\mathbf{r}=(x,y,z)$ and time $t$.  Considering a passive receiver, $\bar{y}_i(t)$ is obtained as
\begin{IEEEeqnarray}{lll} \label{Eq:CR}
\bar{y}_i(t) = \underset{\mathbf{r}\in\mathcal{V^{\mathrm{rx}}}}{\iiint} C_i(\mathbf{r},t) \mathrm{d}\mathbf{r},\quad i\in\{A,B\},
\end{IEEEeqnarray}
where $\mathcal{V^{\mathrm{rx}}}$ is the set of points within the volume of the receiver.  Concentrations $C_A(\mathbf{r},t)$ and  $C_B(\mathbf{r},t)$  can be found using the following reaction-diffusion equations 
\cite{Nariman_Acid,ReactionDiffSim}
\begin{IEEEeqnarray}{lll} \label{Eq:Reaction_Diff}
\frac{\partial C_A(\mathbf{r},t)}{\partial t} = 
D_A \nabla^2 C_A(\mathbf{r},t) - k_f C_A(\mathbf{r},t)C_B(\mathbf{r},t) + k_b \quad \IEEEyesnumber\IEEEyessubnumber \\
\frac{\partial C_B(\mathbf{r},t)}{\partial t} = 
D_B \nabla^2 C_B(\mathbf{r},t) - k_f C_A(\mathbf{r},t)C_B(\mathbf{r},t) + k_b, \quad\,\,\, \IEEEyessubnumber 
\end{IEEEeqnarray}
where $\nabla^2 = \frac{\partial^2 }{\partial x^2} + \frac{\partial^2 }{\partial y^2} + \frac{\partial^2 }{\partial z^2}$. As stated earlier, the general reaction-diffusion equations in (\ref{Eq:Reaction_Diff}) have not yet been solved in closed form. The difficulty mainly arises from  the coupling of the two equations and the non-linear term $ k_f C_A(\mathbf{r},t)C_B(\mathbf{r},t)$. Note that even if we assume that one of the variables, e.g. $C_B(\mathbf{r},t)$, is fixed, it is still challenging to solve (\ref{Eq:Reaction_Diff}a) in terms of $C_A(\mathbf{r},t)$.  Therefore, in the following, we derive a numerical method to solve (\ref{Eq:Reaction_Diff}) in a computationally efficient manner. This is achieved by fully exploiting the simplifying characteristics of our system model. 

\subsection{Derivation of the CR}\label{Sec:CR}

Let us assume that time is divided into a series of small intervals of length $\Delta t$. The main idea behind the proposed approach for computing the CR is to find the concentrations at the end of each time interval given the concentrations at the beginning of the time interval  while exploiting the condition $\Delta t\to 0$. In particular, from the reaction diffusion equations in (\ref{Eq:Reaction_Diff}), we have
\begin{IEEEeqnarray}{lll} \label{Eq:Reaction_Diff_integral}
C_i(\mathbf{r},t+\Delta t)= 
C_i(\mathbf{r},t) + G_i(\mathbf{r},t)  \nonumber \\
+\underset{\Delta C_i^{\mathrm{diff}}(\mathbf{r},t)}{\underbrace{\int_{\tilde{t}=t}^{t+\Delta t} \hspace{-6mm} D_i \nabla^2 C_i(\mathbf{r},\tilde{t}) \mathrm{d}\tilde{t}}}  
+\hspace{-2mm} \underset{\Delta C_i^{\mathrm{react}}(\mathbf{r},t)}{\underbrace{\int_{\tilde{t}=t}^{t+\Delta t} \hspace{-5mm} \left(- k_f C_A(\mathbf{r},\tilde{t})C_B(\mathbf{r},\tilde{t}) + k_b\right) \hspace{-1mm} \mathrm{d}\tilde{t}}},\,\,\,
\quad
\end{IEEEeqnarray}
where $G_i(\mathbf{r},t)= \sum_{t_i\in\mathcal{T}_i} N_i^{\mathrm{tx}} \delta(\mathbf{r},t - t_i)$ represents the concentration of type-$i$ molecules that are released by the transmitter into the channel where $\delta(\mathbf{r},t)=\delta(x)\delta(y)\delta(z)\delta(t)$ and $\delta(\cdot)$ is the Dirac delta function. Moreover, $\Delta C_i^{\mathrm{diff}}(\mathbf{r},t)$ and $\Delta C_i^{\mathrm{react}}(\mathbf{r},t)$ are concentration changes due to  diffusion and reaction, respectively.  The following proposition specifies $C_i(\mathbf{r},t+\Delta t)$ for $\Delta t\to 0$.

\begin{prop}\label{Prop:Numerical}
For the MC system under consideration, assuming $\Delta t\to 0$ and that the release times $t\in\mathcal{T}_i$ are integer multiplies of $\Delta t$, we obtain 
\begin{IEEEeqnarray}{lll} \label{Eq:React_Diff_Solution}
C_i(\mathbf{r},t+\Delta t)\, & = 
G_i(\mathbf{r},t+\Delta t) 
 +\bar{C}_i^{\mathrm{diff}}(\mathbf{r},t)
 +\Delta \bar{C}_i^{\mathrm{react}}(\mathbf{r},t), \quad\,\,
\end{IEEEeqnarray}
where
\begin{IEEEeqnarray}{rll}
\bar{C}_i^{\mathrm{diff}}(\mathbf{r},t)  &=    
 \frac{1}{(4\pi D_i \Delta t)^{\frac{3}{2}}} 
\iiint_{\tilde{\mathbf{r}}}
C_i(\tilde{\mathbf{r}},t) e^{-\frac{\|\mathbf{r}-\tilde{\mathbf{r}}\|^2}{4D_i \Delta t}}\mathrm{d}\tilde{\mathbf{r}}\label{Eq:Diff_Solution}  \\
\Delta \bar{C}_i^{\mathrm{react}}(\mathbf{r},t) &=-\big(k_fC_A(\mathbf{r},t)C_B(\mathbf{r},t)+ k_b\big)\Delta t. \label{Eq:Reaction_Solution}
\end{IEEEeqnarray}
\end{prop}
\begin{IEEEproof}
The proof is given in Appendix~\ref{App:Prop_Numerical}.
\end{IEEEproof}

 Note that the key assumptions that we made for (\ref{Eq:React_Diff_Solution}) to hold are the \textit{unbounded environment} and the \textit{passive receiver} such that  no boundary conditions are imposed. The main computational complexity originates from (\ref{Eq:Diff_Solution}) since for each update, a three-dimensional integral has to be evaluated for each point of space $\mathbf{r}\in\mathbb{R}^3$ where $\mathbb{R}$ is the set of real numbers.  Nevertheless, for the commonly adopted assumption of a point-source transmitter with impulsive release \cite{Adam_Enzyme,Nariman_Acid,Reza_Reaction},  the computation of $C_{A}(\mathbf{r},t)$ and $C_{B}(\mathbf{r},t)$ can be significantly simplified using the following corollary.

\begin{corol}\label{Corol:OneD}
Assuming impulsive release from a point-source transmitter located at the origin of the Cartesian coordinates, an unbounded environment, and a passive receiver, the concentrations of the molecules are  only functions of variable $r\triangleq \|\mathbf{r}\|$. In this case, assuming $\Delta t\to 0$, we obtain
\begin{IEEEeqnarray}{rll} \label{Eq:Concen_OneD}
C_{i}(r,t+\Delta t) \,
&= \sum_{t_i\in\mathcal{T}_i}N_i^{\mathrm{tx}}\delta(r,t+\Delta t-t_i)
\nonumber \\
&\,\,\,\,-k_fC_A(r,t)C_B(r,t)\Delta t + k_b\Delta t \nonumber \\
&\,\,\,\,+ \frac{1}{\sqrt{4\pi D_i \Delta t}} 
\int_{\tilde{r}=0}^{\infty} C_i(\tilde{r},t) 
W_i(r,\tilde{r}) \mathrm{d}\tilde{r},\quad
\end{IEEEeqnarray}
where $W_i(r,\tilde{r})$ is given by
\begin{IEEEeqnarray}{lll} \label{Eq:WeightFunc}
W_i(r,\tilde{r}) &=  \frac{2\tilde{r}}{r} 
\exp\left(-\frac{\tilde{r}^2+r^2}{4D_i \Delta t}\right)
\sinh\left(\frac{r\tilde{r}}{2D_i\Delta t}\right). 
\end{IEEEeqnarray}
\end{corol}
\begin{IEEEproof}
The proof is given in Appendix~\ref{App:Corol_OneD}.
\end{IEEEproof}

Note that the three-dimensional integral in (\ref{Eq:Diff_Solution}) is simplified to a one-dimensional integral in the last term of (\ref{Eq:Concen_OneD}) which has to be evaluated for a one-dimensional space variable $r\in\mathbb{R}$. In addition, the term $W_i(r,\tilde{r})$ in the integral  does not depend on the concentrations. Hence, we can evaluate it offline and use it for online concentration updates. In other words, the integral in (\ref{Eq:Concen_OneD}) simplifies to summation and multiplication operations.  Algorithm~\ref{Alg:CR} summarizes the simulation steps for CR computation using Corollary~\ref{Corol:OneD}.

\begin{algorithm}[t]
\caption{Computation of CR}
 \begin{algorithmic}[1]\label{Alg:CR}
 \STATE \textbf{initialize:} $t=0$, $\Delta t$, $T^{\max}$, $\mathcal{T}_i$, and $C_{i}(\mathbf{r},0)$.   
      \WHILE{$t\leq T^{\max}$}
              \STATE Update $t$ with  $t+\Delta t$.
        \STATE Compute $C_{A}(\mathbf{r},t)$ and $C_{B}(\mathbf{r},t)$ from (\ref{Eq:Concen_OneD}).
      \ENDWHILE
      \STATE Return $\bar{y}_A(t)$ and $\bar{y}_B(t)$ from (\ref{Eq:CR}) as the CR.
  \end{algorithmic}
\end{algorithm}


\subsection{Discussion on Complexity}\label{Sec:Complexity}

In the following, we compare the computational complexity of the proposed algorithm and the statistical model with the conventional numerical methods and particle-based simulation.

\subsubsection{Conventional Numerical Methods} Most numerical methods in the literature rely on discretization of space and time to solve the reaction-diffusion equations in (\ref{Eq:Reaction_Diff}) \cite{PDE_numerical,Nariman_Acid}. For instance, for FDM, time  and space are discretized into small intervals to approximate the differential operators in (\ref{Eq:Reaction_Diff}). The advantage of this approach is its universality as it can be applied to different PDEs. However, for the approximations of the differential operators to be accurate, the adopted step size should be very small \cite{PDE_numerical}. Compared to these methods, the proposed approach in Algorithm~\ref{Alg:CR} is a hybrid method where time is discretized; however, the problem is solved analytically with respect to the space variables. Moreover,  the proposed approach is not as sensitive  to the size of the time step length $\Delta t$ as FDM since we do not approximate any differential operator.  Therefore,  the proposed method is much faster than pure numerical methods such as FDM. 

\subsubsection{Particle-based Simulation} Another approach to obtain the CR is to employ stochastic reaction-diffusion simulations to evaluate different realizations of the concentrations of the molecules at each point in space and time and average the resulting concentrations to obtain the expected concentrations. In addition, these realizations can be used to determine the statistics of the concentrations \cite{ReactionDiffSim,Chou_MasterEq,Adam_AcCoRD}. In Section~\ref{Sec:SimResult}, we show the \textit{exact} statistics of the number of molecules observed at the receiver obtained via particle-based simulation. A corresponding simulator is developed for the MC system considered in this paper and is explained in detail in Appendix~\ref{App:SimParticle}.  Similar to the general numerical methods, the advantage of particle-based simulation is its universality as it can be also applied to different MC systems. However, it is very inefficient in terms of computational complexity, especially since for typical MC systems, the number of molecules whose positions have to be tracked in a particle-based simulator can be extremely large. Moreover, the particle-based simulator has to be run many times in order to obtain a sufficient number of samples to develop a statistical model. On the contrary, the complexity of Algorithm~\ref{Alg:CR} does not scale with the number of molecules. Moreover, we show in Section~\ref{Sec:SimResult} that the Poisson distribution with a mean, which is a non-linear function of the ISI-causing sequence  and obtained with Algorithm~\ref{Alg:CR}, can accurately model the statistics of the number of molecules observed at the receiver.

\section{Simulation Results}\label{Sec:SimResult}

In this section, we evaluate the accuracy of the proposed CR computation algorithm and the Poisson statistical model with non-linear ISI and determine the performance of the proposed detectors. Unless stated otherwise, the default values for the system parameters are given in Table~I.  The time step size is  $\Delta t=1\,\mu$s in Algorithm~\ref{Alg:CR} and the particle-based simulator in Appendix~\ref{App:SimParticle}, respectively. Moreover, the simulation results shown in Figs.~\ref{Fig:CR_nonLinear} and \ref{Fig:Poisson_PDF} are obtained by  running the particle-based simulator in Appendix~\ref{App:SimParticle} $10^4$ times for time interval $[0,T^{\max}]$ where $T^{\max}=60\,\mu$s.

\begin{table}
\label{Table:Parameter}
\caption{Default Values of the System Parameters \cite{Adam_Enzyme,Nariman_Acid}. \vspace{-0.2cm}} 
\begin{center}
\scalebox{0.6}
{
\begin{tabular}{|| c | c | c ||}
  \hline 
  Variable & Definition & Value \\ \hline \hline
       $N^{\mathrm{tx}}_A,N^{\mathrm{tx}}_B$ & Number of released molecules  & $5\times 10^3$ molecules \\ \hline      
        $d$ &  Distance between  transmitter and  receiver  & $250$ nm\\ \hline 
         $D_A,D_B$ &  Diffusion coefficient& $ 10^{-9}$ $\text{m}^2\cdot\text{s}^{-1}$\\ \hline          
       $k_f$ &  Forward reaction rate  & $10^{-17}$  molecule$^{-1}$m$^3$s$^{-1}$\\   \hline
       $k_b$ &  Backward reaction rate  & $10^{17}$  molecule$^{-1}$m$^3$s$^{-1}$\\   \hline
              $r$ & Receiver radius   & $50$ nm \\    \hline 
       $T^{\mathrm{symb}}$ &  Symbol duration  & $20$  $\mu$s\\   \hline
\end{tabular}
}
\end{center}\vspace{-0.6cm}
\end{table}

In Fig.~\ref{Fig:CR_nonLinear}, we plot the \textit{expected} number of molecules observed at the receiver, i.e., CR, for three consecutive symbol intervals and data sequence  $[0,1,0]$. For non-reactive signaling molecules, we show analytical results from \cite[Eq.~(1) and (2)]{TCOM_MC_CSI},  and for reactive signaling molecules, we show numerical results obtained with Algorithm~\ref{Alg:CR}  and simulation results generated with the particle-based simulator introduced in Appendix~\ref{App:SimParticle}. From Fig.~\ref{Fig:CR_nonLinear}, we observe a perfect agreement between the numerical and simulation results. For the first symbol, where  ISI does not exist, the peak concentration of type-$A$ molecules is identical for both reactive and non-reactive signaling whereas the contribution of ISI observed in the next symbol interval is much higher for the non-reactive case compared to reactive signaling.  Moreover, as can be seen from Fig.~\ref{Fig:CR_nonLinear},  for non-reactive signaling, the concentration of the molecules increases over time. On the contrary, for reactive signaling, the peak concentration of the received molecules remains almost constant. This is due to the fact that signaling molecules participate in a degradation reaction and cancel each other out. Although  we assumed an unbounded simulation environment here, we expect that degradation via reactive molecules becomes even more important in a bounded environment where the accumulation of molecules can significantly contaminate the channel. 

\begin{figure} 
  \centering\vspace{-0.1cm}
\hspace{-0.6cm}
\resizebox{1\linewidth}{!}{\psfragfig{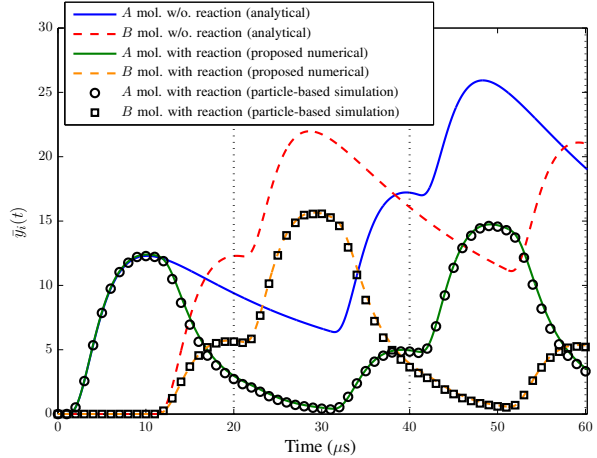}} \vspace{-0.4cm}
\caption{CR, $\bar{y}_A(t)$ and $\bar{y}_B(t)$, for sequence $[0,1,0]$. The dotted line shows the start of a new symbol interval. \vspace{-0.5cm} }
\label{Fig:CR_nonLinear}
\end{figure}

Next, we evaluate the accuracy of the statistical model introduced in (\ref{Eq:Poisson_ISI}). We assume that the sampling time is $t_s=\tau_A=\tau_B$. We choose  again data sequence $[0,1,0]$ whose corresponding CR is plotted in Fig.~\ref{Fig:CR_nonLinear}.  Fig.~\ref{Fig:Poisson_PDF} shows  the histogram of the number of molecules observed at the receiver, which is obtained via particle-based simulation, i.e., the ``exact distribution", and the proposed Poisson model in (\ref{Eq:Poisson_ISI}). The blue, red, and green curves correspond to the observations in the first, second, and third symbol interval, respectively. Note that from Fig.~\ref{Fig:CR_nonLinear}, we observe that $y_A$ has a small mean (less than $5$) in the second symbol interval and a large mean (more than $10$) in the first and third symbol intervals. Similarly,  $y_B$ has a small mean in the first and third symbol intervals and a large mean in the second symbol interval.   Fig.~\ref{Fig:Poisson_PDF} includes two subplots that focus on the small means (left-hand side subplot) and the large means (right-hand side subplot). As can be seen from Fig.~\ref{Fig:Poisson_PDF}, the Poisson model with non-linear ISI can accurately model the histograms obtained from particle-based simulation for both small and large concentration means. This validates the Poisson model for MC systems with reactive signaling and is in line with the results reported in the chemistry and physics literature \cite{PoissonGardiner,CoxNatureCommun}. 

\begin{figure} 
  \centering\vspace{-0.1cm}
\hspace{-0.6cm}
\resizebox{1\linewidth}{!}{\psfragfig{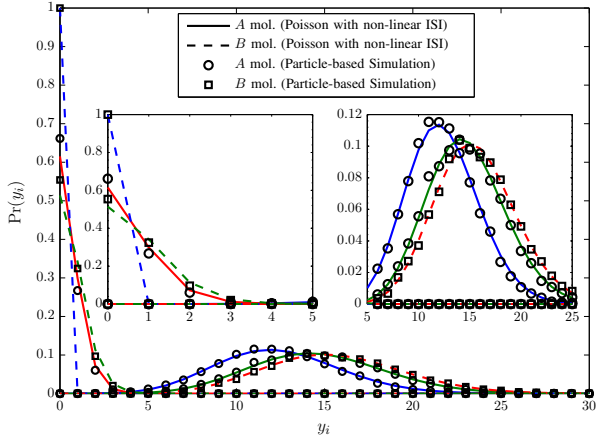}} \vspace{-0.4cm}
\caption{PMFs of observation $y_A$ and $y_B$ for sequence $[0,1,0]$ and MCs with reactive signaling molecules. The blue, red, and green curves correspond to the observations in the first, second, and third symbol interval, respectively. \vspace{-0.5cm} }
\label{Fig:Poisson_PDF}
\end{figure}

Finally, we compare the performance of the reactive MC system proposed in this work with that of MC systems with no chemical reactions. Particularly, the following two MC systems with non-reactive molecules are considered. The first system uses the same modulation scheme as introduced in Section~\ref{Sec:Transmitter} but with {\em no chemical reactions}. This was originally proposed in \cite{Gau_AB} where the benefit of employing two types of molecules is the resulting diversity. Second, we consider a system model with only a single type of molecule and on-off keying (OOK) modulation. This system has been used in many previous works \cite{Nariman_Survey}, and here we use the detection algorithm developed in \cite[Eq.~(6)]{NanoCOM16} for calculating the BER. For the simulation results, we consider $10^5$ realizations of blocks of $10$ symbols and use $L=3$ for the proposed ML detector. We emphasize that for simulation of $y_A$ and $y_B$, the memory of all previous symbol intervals is considered whereas for the ML detector, only a memory length of size three is assumed for simplicity. 

\begin{figure} 
  \centering\vspace{-0.1cm}
\hspace{-0.6cm}
\resizebox{1\linewidth}{!}{\psfragfig{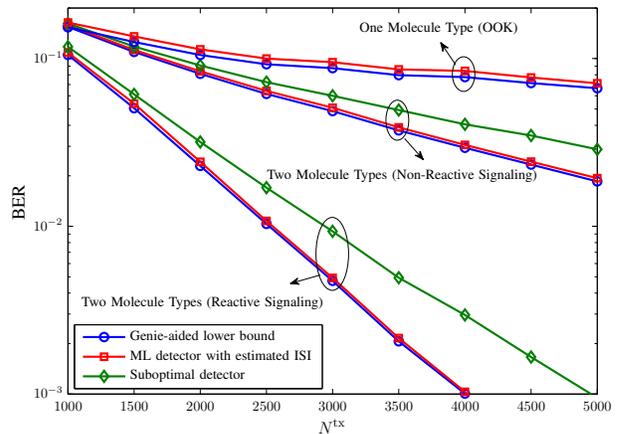}} \vspace{-0.4cm}
\caption{BER vs. number of released molecules, $N^{\mathrm{tx}}_A = N^{\mathrm{tx}}_B\triangleq N^{\mathrm{tx}}$. \vspace{-0.5cm} }
\label{Fig:BER}
\end{figure}

Fig.~\ref{Fig:BER} shows the BER vs. the number of released molecules, $N^{\mathrm{tx}}_A = N^{\mathrm{tx}}_B\triangleq N^{\mathrm{tx}}$, for the genie-aided lower bound from Proposition~\ref{Prop:MLDetector}, the ML detector with estimated ISI, and the suboptimal detector in (\ref{Eq:MLdetector_simple}). We observe from Fig.~\ref{Fig:BER} that the MC system with reactive signaling molecules has a superior performance compared to the MC system with non-reactive signaling molecules for all considered detectors.  This is due to the fact that  for the adopted symbol duration, ISI is sufficiently reduced with reactive signaling whereas for non-reactive signaling, the ISI is severe, cf. Fig.~\ref{Fig:CR_nonLinear}. Moreover, even non-reactive signaling with two types of molecules outperforms OOK signaling with one type of molecule because of  the  diversity gain that observing two types of molecules provides. Furthermore, Fig.~\ref{Fig:BER} shows that the ML detector with estimated ISI  performs very close to the genie-aided lower bound. Finally, the proposed suboptimal detector in (\ref{Eq:MLdetector_simple}) performs well without requiring knowledge of the CR which makes it suitable for MC systems with limited computational capabilities.

\section{Conclusions}\label{Sec:Conclusions}

We studied a binary MC system with reactive signaling molecules and  derived  an optimal genie-aided ML detector and two suboptimal detectors for this system. Since the CR was needed for the genie-aided ML detector and one of the suboptimal detectors as well as for the performance evaluation of all three detectors,  we  developed a numerical algorithm for efficient CR computation. The accuracy of this algorithm has been validated via particle-based simulations.   Moreover, simulation results showed the superiority in performance of reactive systems over non-reactive systems  due to efficient ISI reduction and revealed the similar performance of the proposed suboptimal detectors compared to the genie-aided detector. As part of future work, we will evaluate the accuracy of this model and the performance of the proposed detection algorithms with an experimental platform.

\appendices

\section{}\label{App:SimParticle} 

For particle-based simulation, the positions of individual particles are tracked during the simulation time. In the following, we explain how we update the position of the molecules for the considered release, diffusion, and reaction mechanisms.

\subsubsection{Transmitter Release} Let $\mathbf{r}_i(t)=(x_i(t),y_i(t),z_i(t))$ denote the coordinate of a specific type-$i$ molecule at time instance $t$.  For instantaneous release from a point-source transmitter, we simply place $N^{\mathrm{tx}}_i$ type-$i$ molecules at position $\mathbf{r}_i(t) = (0,0,0)$ at any release time instant $t\in\mathcal{T}_i$. 

\subsubsection{Diffusion} 
According to Brownian dynamics, the position of each molecule at time instance $t+\Delta t$ is updated as \cite{ReactionDiffSim}
\begin{IEEEeqnarray}{lll} \label{Eq:Brownian}
\mathbf{r}_i(t+\Delta t) = \mathbf{r}_i(t) + \sqrt{2D_i\Delta t} \big(\Delta x_i,\Delta y_i,\Delta z_i),
\end{IEEEeqnarray}
where $\Delta x_i,\Delta y_i,\Delta z_i\sim\mathcal{N}(0,1)$ and $\mathcal{N}(\mu,\sigma^2)$ denotes a normal RV with mean $\mu$ and variance $\sigma^2$.

\subsubsection{Forward Reaction}

The forward reaction in (\ref{Eq:Reaction}), i.e., $ A+B\overset{k_f}{\rightarrow} \varnothing$, is a second order bimolecular reaction.
The fundamental rule for stochastic simulation of these reactions is based on the fact that a reaction occurs within interval $[t, t+\Delta t]$ when the reactant molecules are within a certain binding radius $\rho_b$ \cite{ReactionDiffSim}.  Therefore, in our simulation, we can simply remove one type-$A$ and one type-$B$ molecule if their distance is less than $\rho_b$. Unfortunately, the exact value of $\rho_b$ cannot be found analytically in general  and depends on reaction rate $k_f$ and the choice of $\Delta t$. Nevertheless, for the two special cases of $\Delta t\to 0$ and $\Delta t\to \infty$, the following simple relations are available \cite[Eqs. (19) and (20)]{ReactionDiffSim}
\begin{IEEEeqnarray}{lll} \label{Eq:BindingRadius}
k_f = \begin{cases}
4\pi\rho_b(D_A+D_B),\quad &\mathrm{if}\,\,\rho_{\mathrm{rms}}\ll \rho_b \\
4\pi\rho_b^3/(3\Delta t),\quad &\mathrm{if}\,\,\rho_{\mathrm{rms}} \gg \rho_b
\end{cases}
\end{IEEEeqnarray}
where $\rho_{\mathrm{rms}}=\sqrt{2(D_A+D_B)\Delta t}$ is the mutual root mean square step length of type-$A$ and type-$B$ molecules. Using the simplified formula for $\rho_b$ in (\ref{Eq:BindingRadius}), we obtain equivalent conditions for $\rho_{\mathrm{rms}} \ll \rho_b$ and $\rho_{\mathrm{rms}} \gg \rho_b$ as
\begin{IEEEeqnarray}{lll} \label{Eq:BindingRadius_Time}
\begin{cases}
\rho_{\mathrm{rms}}\ll \rho_b \rightarrow \Delta t \ll \frac{k_f^2}{32\pi^2(D_A+D_B)^3}\\
\rho_{\mathrm{rms}} \gg \rho_b \rightarrow \Delta t \gg \frac{9k_f^2}{128\pi^2(D_A+D_B)^3}
\end{cases}
\end{IEEEeqnarray}
In order to reduce the computational complexity, we choose $\Delta t$ sufficiently large such that condition $\rho_{\mathrm{rms}} \gg \rho_b$ holds.

\subsubsection{Backward Reaction}
The backward reaction in  (\ref{Eq:Reaction}) is in form of  zeroth order reaction $\varnothing \overset{k_b}{\rightarrow} A + B$. Suppose that the simulation environment is a cube of volume $V= L^3$. Moreover, let RV $n(t)$ denote the number of times that the backward reaction occurs in a time interval $[t,t+\Delta t]$. Then, $n(t)$ follows a Poisson distribution \cite{ReactionDiffSim}
\begin{IEEEeqnarray}{lll} \label{Eq:ZeroOrderPoisson}
n(t) = \mathcal{P}\big(Vk_b\Delta t\big).
\end{IEEEeqnarray}
Here, we have to be careful about the positions of  the type-$A$ and type-$B$ molecules that are generated via the backward reaction. In particular, if $\rho_{\mathrm{rms}} \ll \rho_b$ holds and we put these molecules on the same location, then these molecules directly participate in the forward reaction before they can diffuse away  regardless of the value of $\rho_b$.  In order to avoid the automatic degradation of  type-$A$ and type-$B$ molecules, the type-$A$ and type-$B$ product molecules are initially separated by a fixed distance which is larger than $\rho_b$ denoted by the unbinding radius $\rho_u$.
 Let  $\mathbf{l}$ denote the center of the cube of the simulation environment. Then,  the zeroth order reaction can be simulated by inserting each of the $n(t)$ molecules of type-$i$ at random positions $\mathbf{r}_i(t)$  obtained as \cite{ReactionDiffSim}
\begin{IEEEeqnarray}{lll} \label{Eq:ZeroOrderPosition}
\mathbf{r}_A(t) &= \mathbf{l} +  L\big(\Delta x_A,\Delta y_A,\Delta z_A\big), \IEEEyesnumber\IEEEyessubnumber \\
\mathbf{r}_B(t) &= \mathbf{r}_A(t) + \rho_u\big(\Delta x_B,\Delta y_B,\Delta z_B\big),\IEEEyessubnumber
\end{IEEEeqnarray}
where $\Delta x_A,\Delta y_A,\Delta z_A\sim\mathcal{U}(-0.5,0.5)$ and $\mathcal{U}(a,b)$ is an RV uniformly distributed in interval $[a,b]$. Moreover, $\Delta x_B,\Delta y_B$, and $\Delta z_B$ are any numbers satisfying $\Delta x_B^2+\Delta y_B^2+\Delta z_B^2=1$. On the other hand, if $\rho_{\mathrm{rms}} \gg \rho_b$ holds and diffusion is simulated after the backward reaction in the adopted simulator, it is very likely that the molecules diffuse out of the binding radius after one diffusion step. In this case, the value of the unbinding is not important and  without loss of generality, we can choose $\rho_u=0$ which leads to $\mathbf{r}_A(t) = \mathbf{r}_B(t)$.

Algorithm~\ref{Alg:Particle} summarizes the main steps required for particle-based simulation of the considered MC system.

\begin{algorithm}[t]
\caption{Particle-based Simulation}
 \begin{algorithmic}[1]\label{Alg:Particle}
 \STATE \textbf{initialize:} $t=0$, $\Delta t$, $T^{\max}$, and $\mathcal{T}_i$.   
      \WHILE{$t\leq T^{\max}$}
      \IF{$t\in\mathcal{T}_i$}
      \STATE \textit{Transmitter input:}  Add $N^{\mathrm{tx}}_i$ type-$i$ molecules at position $\mathbf{r}_i(t) = (0,0,0)$.
      \ENDIF
       \STATE \textit{Diffusion:} Update the positions of molecules $\mathbf{r}_i(t),\,\,i\in\{A,B\}$, based on~(\ref{Eq:Brownian}).
       \STATE \textit{Degradation:} Remove pairs of type-$A$ and type-$B$ molecules whose positions satisfy $\|\mathbf{r}_A-\mathbf{r}_B\|\leq\rho_b$.
       \STATE \textit{Production:} Add $n(t)$ pairs of type-$A$ and type-$B$ molecules at positions $\mathbf{r}_A(t)$ and $\mathbf{r}_B(t)$, respectively,  given in (\ref{Eq:ZeroOrderPosition}).
       \STATE Assign $y_i(t),\,\,i\in\{A,B\}$, as the number  of type-$i$ molecules whose positions satisfy $\mathbf{r}_i\in\mathcal{V}^{\mathrm{rx}}$.
              \STATE Update $t$ with  $t+\Delta t$.
      \ENDWHILE
      \STATE Return $y_A(t)$ and $y_B(t)$.
  \end{algorithmic}
\end{algorithm}

\section{}\label{App:Prop_MLDetector} 

The log likelihood ratio (LLR) for  problem (\ref{Eq:MLprob})  can be written as
\begin{IEEEeqnarray}{lll} \label{Eq:LLR}
\text{LLR}&=\log\left(\frac{f_{\mathcal{P}}(y_A|s=0,\mathbf{s})f_{\mathcal{P}}(y_B|s=0,\mathbf{s})}{f_{\mathcal{P}}(y_A|s=1,\mathbf{s})f_{\mathcal{P}}(y_B|s=1,\mathbf{s})}\right) \nonumber \\
&= \log\left(\frac{(\bar{y}_A^{(0)}(\mathbf{s}))^{y_A} e^{-\bar{y}_A^{(0)}(\mathbf{s})} (\bar{y}_B^{(0)}(\mathbf{s}))^{y_B} e^{-\bar{y}_B^{(0)}(\mathbf{s})}}{(\bar{y}_A^{(1)}(\mathbf{s}))^{y_A} e^{-\bar{y}_A^{(1)}(\mathbf{s})} (\bar{y}_B^{(1)}(\mathbf{s}))^{y_B} e^{-\bar{y}_B^{(1)}(\mathbf{s})}}\right) \nonumber \\
&= y_A \log\left(\frac{\bar{y}_A^{(0)}(\mathbf{s})}{\bar{y}_A^{(1)}(\mathbf{s})}\right)
-y_B \log\left(\frac{\bar{y}_B^{(1)}(\mathbf{s})}{\bar{y}_B^{(0)}(\mathbf{s})}\right)\nonumber \\
&\quad -\bar{y}_A^{(0)}(\mathbf{s})-\bar{y}_B^{(0)}(\mathbf{s})+\bar{y}_A^{(1)}(\mathbf{s})+\bar{y}_B^{(1)}(\mathbf{s}).
\end{IEEEeqnarray}
Due to the monotonicity of the logarithm, the ML problem in (\ref{Eq:MLprob}) can be rewritten as $\text{LLR} \overset{s=0}{\underset{s=1}{\gtreqless}} 0$.  Defining $\alpha(\mathbf{s})$ and $\beta(\mathbf{s})$  as in Proposition~\ref{Prop:MLDetector}, we obtain (\ref{Eq:MLdetector}) which concludes the proof.

\section{}\label{App:Prop_Numerical}

In the following, we discuss the concentration changes due the reaction and diffusion processes.
\subsubsection{Reactions}
Assuming $\Delta t\to 0$, the concentration $C_i(\mathbf{r},\tilde{t}),\,\,\tilde{t}\in(t,t+\Delta t]$, can be written as 
\begin{IEEEeqnarray}{lll} 
C_i(\mathbf{r},\tilde{t})=C_i(\mathbf{r},t)+\epsilon_i(\mathbf{r},\tilde{t}), 
\quad \tilde{t}\in(t,t+\Delta t],
\end{IEEEeqnarray}
where $\epsilon_i(\mathbf{r},\tilde{t})$ is the concentration change due to the reaction and diffusion processes at any time $\tilde{t}$ within the interval $(t,t+\Delta t]$.  Let us define $\epsilon^{\max}_i(\mathbf{r},t)=\max_{\tilde{t}} \epsilon_i(\mathbf{r},\tilde{t}),\,\,\tilde{t}\in(t,t+\Delta t]$, and $\epsilon^{\min}_i(\mathbf{r},t)=\min_{\tilde{t}} \epsilon_i(\mathbf{r},\tilde{t}),\,\,\tilde{t}\in(t,t+\Delta t]$. 
Assuming $\Delta t\to 0$, $\Delta C_i^{\mathrm{react}}(\mathbf{r},t)$ defined in (\ref{Eq:Reaction_Diff_integral}) is bounded as
\begin{IEEEeqnarray}{lll} \label{Eq:Reaction_Bound_React}
\Delta C_i^{\mathrm{react}}(\mathbf{r},t) 
&\leq \Delta \bar{C}_i^{\mathrm{react}}(\mathbf{r},t) 
+ \epsilon^{\max}_A(\mathbf{r},t)C_B(\mathbf{r},t)\Delta t
\nonumber \\
&\,\,+ \epsilon^{\max}_B(\mathbf{r},t)C_A(\mathbf{r},t)\Delta t
+ \epsilon^{\max}_A(\mathbf{r},t)
\epsilon^{\max}_B(\mathbf{r},t) \Delta t  \nonumber\\
&\overset{(a)}{=} \Delta \bar{C}_i^{\mathrm{react}}(\mathbf{r},t)+ o(\Delta t) \IEEEyesnumber\IEEEyessubnumber\\
\Delta C_i^{\mathrm{react}}(\mathbf{r},t) 
&\geq \Delta \bar{C}_i^{\mathrm{react}}(\mathbf{r},t) 
+ \epsilon^{\min}_A(\mathbf{r},t)C_B(\mathbf{r},t)\Delta t
\nonumber \\
&\,\,+ \epsilon^{\min}_B(\mathbf{r},t)C_A(\mathbf{r},t)\Delta t
+ \epsilon^{\min}_A(\mathbf{r},t)
\epsilon^{\min}_B(\mathbf{r},t) \Delta t  \nonumber\\
&\overset{(b)}{=} \Delta \bar{C}_i^{\mathrm{react}}(\mathbf{r},t)+ o(\Delta t),
\IEEEyessubnumber
\end{IEEEeqnarray}
where  $o(\cdot)$ represents the little-$o$ notation,  $\Delta \bar{C}_i^{\mathrm{react}}(\mathbf{r},t)$ is given by (\ref{Eq:Reaction_Solution}), and  we used $\epsilon^{\max}_i(\mathbf{r},t)\to 0$ and $\epsilon^{\min}_i(\mathbf{r},t)\to 0$ as $\Delta t\to 0$ for equalities $(a)$ and $(b)$, respectively. 
Therefore, the concentration $\Delta C_i^{\mathrm{react}}(\mathbf{r},t)$ can be approximated by
$\Delta\bar{C}_i^{\mathrm{react}}(\mathbf{r},t)$  assuming $\Delta t \to 0$. 

\subsubsection{Diffusion} With a similar argument as  for the approximation of $\Delta C_i^{\mathrm{react}}(\mathbf{r},t)$, we can show that $\Delta C_i^{\mathrm{diff}}(\mathbf{r},t)\to D_i \nabla^2 C_i(\mathbf{r},t) \Delta t$ as $\Delta t\to 0$ if $\nabla^2 C_i(\mathbf{r},t)$ is non-zero\footnote{Note that if $\nabla^2 C_i(\mathbf{r},t)$ is zero, the overall impact of $\Delta C_i^{\mathrm{diff}}(\mathbf{r},t)$ is negligible compared to $\Delta C_i^{\mathrm{react}}(\mathbf{r},t)$.}. In other words, $\Delta C_i^{\mathrm{diff}}(\mathbf{r},t)$ is on the order of $\Delta t$. Using this result, in the following, we provide an alternative approximation of $\Delta C_i^{\mathrm{diff}}(\mathbf{r},t)$ which does not involve the Laplace  operator $\nabla^2$. Assuming $\Delta t\to 0$, the concentration $C_i(\mathbf{r},\tilde{t}),\,\,\tilde{t}\in(t,t+\Delta t]$, can be written as 
\begin{IEEEeqnarray}{lll} \label{Eq:Reaction_Bound_Diff}
C_i(\mathbf{r},\tilde{t})=C^{\mathrm{diff}}_i(\mathbf{r},\tilde{t})+\delta_i(\mathbf{r},\tilde{t}), 
\quad \tilde{t}\in(t,t+\Delta t],
\end{IEEEeqnarray}
where $C^{\mathrm{diff}}_i(\mathbf{r},\tilde{t})$ is the concentration assuming reaction does not occur within $(t,t+\Delta t]$ and  $\delta_i(\mathbf{r},\tilde{t})$ models the concentration difference due to reaction processes. Assuming $\Delta t \to 0$,  $\Delta C_i^{\mathrm{diff}}(\mathbf{r},t)$ is given  by
\begin{IEEEeqnarray}{lll} \label{Eq:Diff_Bound}
\Delta C_i^{\mathrm{diff}}(\mathbf{r},t)  = \Delta \bar{C}_i^{\mathrm{diff}}(\mathbf{r},t) +  o(\Delta t),
\end{IEEEeqnarray}
where  $\Delta \bar{C}_i^{\mathrm{diff}}(\mathbf{r},t)=\bar{C}_i^{\mathrm{diff}}(\mathbf{r},t)-C_i(\mathbf{r},t)$ and $\bar{C}_i^{\mathrm{diff}}(\mathbf{r},t)$ is the concentration due to free diffusion without reaction which is given by (\ref{Eq:Diff_Solution}) \cite[Chapter 1.7]{NonlinearPDE_Debnath}. Since $\Delta C_i^{\mathrm{diff}}(\mathbf{r},t)$ is on the order of $\Delta t$, it can be approximated by  $\bar{C}_i^{\mathrm{diff}}(\mathbf{r},t)$ assuming $\Delta t\to 0$. 

Substituting $\Delta \bar{C}_i^{\mathrm{diff}}(\mathbf{r},t)=\bar{C}_i^{\mathrm{diff}}(\mathbf{r},t)-C_i(\mathbf{r},t)$ and $\Delta \bar{C}_i^{\mathrm{react}}(\mathbf{r},t)$ for $ \Delta C_i^{\mathrm{diff}}(\mathbf{r},t)$ and $\Delta C_i^{\mathrm{react}}(\mathbf{r},t)$ into (\ref{Eq:Reaction_Diff_integral}), respectively, yields the update rule in (\ref{Eq:React_Diff_Solution}) and concludes the proof. 

\section{}\label{App:Corol_OneD} 

Because of the geometrical symmetry of the problem, the concentration for the MC system under consideration is only a function of  $r\triangleq \|\mathbf{r}\|$. 
The simplification for $C_i^{\mathrm{diff}}(\mathbf{r},t+\Delta t)$ follows from transforming an integral from  Cartesian coordinates to spherical coordinates with variables $r\geq 0$, $\phi\in[0,\pi]$, and $\theta\in[0,2\pi]$. Moreover, without loss of generality, we consider $\mathbf{r}=(0,0,r)$ in order to simplify (\ref{Eq:Diff_Solution}) as follows
\begin{IEEEeqnarray}{lll} 
C_i^{\mathrm{diff}}(\mathbf{r},t+\Delta t) \nonumber  \\
 =  \frac{1}{(4\pi D_i \Delta t)^{3/2}} 
\iiint_{\tilde{\mathbf{r}}}
C_i(\tilde{\mathbf{r}},t) \exp\left(-\frac{\|\tilde{\mathbf{r}}\|^2+r^2-2r\tilde{z}}{4D_i \Delta t}\right)\mathrm{d}\tilde{\mathbf{r}}, \nonumber \\
 = \frac{1}{(4\pi D_i \Delta t)^{3/2}} 
\int_{\tilde{r}=0}^{\infty} C_i(\tilde{r},t) 
\tilde{r}^2\exp\left(-\frac{\tilde{r}^2+r^2}{4D_i \Delta t}\right)
\nonumber \\ \qquad\qquad\quad\,\,\times
\int_{\tilde{\phi}=0}^{\pi} \int_{\tilde{\theta}=0}^{2\pi}
\sin(\tilde{\phi})\exp\left(\frac{r\tilde{r}\cos(\tilde{\phi})}{2D_i \Delta t}\right) 
\mathrm{d}\tilde{\theta} \mathrm{d}\tilde{\phi}\mathrm{d}\tilde{r}, \quad\,\,\, \nonumber \\
 = \frac{1}{\sqrt{4\pi D_i \Delta t}} 
\int_{\tilde{r}=0}^{\infty} C_i(\tilde{r},t) \nonumber \\ \qquad\qquad\quad\,\,\times
\frac{2\tilde{r}}{r} 
\exp\left(-\frac{\tilde{r}^2+r^2}{4D_i \Delta t}\right)
\sinh\left(\frac{r\tilde{r}}{2D_i\Delta t}\right) \mathrm{d}\tilde{r}, 
\end{IEEEeqnarray}
where  we used the identity $\int_{x=0}^{\pi} \sin(x)\exp(a\cos(x))\mathrm{d}x = \frac{2\sinh(a)}{a}$. Defining $W_i(r,\tilde{r})$ as in (\ref{Eq:WeightFunc}) leads to  (\ref{Eq:Concen_OneD}) and completes the proof.

\bibliographystyle{IEEEtran}
\bibliography{Ref_27_10_2017}

\begin{thebibliography}{10}
\providecommand{\url}[1]{#1}
\csname url@samestyle\endcsname
\providecommand{\newblock}{\relax}
\providecommand{\bibinfo}[2]{#2}
\providecommand{\BIBentrySTDinterwordspacing}{\spaceskip=0pt\relax}
\providecommand{\BIBentryALTinterwordstretchfactor}{4}
\providecommand{\BIBentryALTinterwordspacing}{\spaceskip=\fontdimen2\font plus
\BIBentryALTinterwordstretchfactor\fontdimen3\font minus
  \fontdimen4\font\relax}
\providecommand{\BIBforeignlanguage}[2]{{%
\expandafter\ifx\csname l@#1\endcsname\relax
\typeout{** WARNING: IEEEtran.bst: No hyphenation pattern has been}%
\typeout{** loaded for the language `#1'. Using the pattern for}%
\typeout{** the default language instead.}%
\else
\language=\csname l@#1\endcsname
\fi
#2}}
\providecommand{\BIBdecl}{\relax}
\BIBdecl

\bibitem{Nariman_Survey}
N.~Farsad, H.~Yilmaz, A.~Eckford, C.~Chae, and W.~Guo, ``{A Comprehensive
  Survey of Recent Advancements in Molecular Communication},'' \emph{IEEE
  Commun. Surveys Tutorials}, vol.~18, no.~3, pp. 1887--1919, third quarter
  2016.

\bibitem{Nakano}
T.~Nakano, Y.~Okaie, and J.-Q. Liu, ``{Channel Model and Capacity Analysis of
  Molecular Communication with Brownian Motion},'' \emph{IEEE Commun. Lett.},
  vol.~16, no.~6, pp. 797--800, Jun. 2012.

\bibitem{CellBio}
B.~Alberts, D.~Bray, K.~Hopkin, A.~Johnson, J.~Lewis, M.~Raff, K.~Roberts, and
  P.~Walter, \emph{{Essential Cell Biology}}.\hskip 1em plus 0.5em minus
  0.4em\relax New York, NY: Garland Science, 4th ed., 2014.

\bibitem{Adam_Enzyme}
A.~Noel, K.~Cheung, and R.~Schober, ``{Improving Receiver Performance of
  Diffusive Molecular Communication with Enzymes},'' \emph{IEEE Trans.
  NanoBiosci.}, vol.~13, no.~1, pp. 31--43, Mar. 2014.

\bibitem{Nariman_Acid}
\BIBentryALTinterwordspacing
N.~Farsad and A.~Goldsmith, ``{A Novel Molecular Communication System Using
  Acids, Bases and Hydrogen Ions},'' \emph{Available online on arXiv}, 2015.
  [Online]. Available: \url{http://arxiv.org/abs/1511.08957}
\BIBentrySTDinterwordspacing

\bibitem{Nariman_AcidBasePlatform}
\BIBentryALTinterwordspacing
N.~Farsad, D.~Pan, and A.~Goldsmith, ``{A Novel Experimental Platform for
  In-Vessel Multi-Chemical Molecular Communications},'' \emph{Accepted for
  presentation in IEEE Globecom}, 2017. [Online]. Available:
  \url{https://arxiv.org/abs/1704.04810}
\BIBentrySTDinterwordspacing

\bibitem{NonlinearPDE_Debnath}
L.~Debnath, \emph{{Nonlinear Partial Differential Equations for Scientists and
  Engineers}}.\hskip 1em plus 0.5em minus 0.4em\relax Springer Science \&
  Business Media, 2011.

\bibitem{PDE_numerical}
J.~Lang, ``{Numerical Solution of Reaction-Diffusion Equations},'' in
  \emph{Scientific Computing in Chemical Engineering}.\hskip 1em plus 0.5em
  minus 0.4em\relax Springer, 1996, pp. 136--141.

\bibitem{PDE_coupled}
L.~Somathilake, ``{Numerical Solutions of Reaction-Diffusion Systems with
  Coupled Diffusion Terms},'' \emph{Ruhuna J. Science}, vol.~4, no.~4, 2012.

\bibitem{Reza_Reaction}
R.~Mosayebi, A.~Gohari, M.~Mirmohseni, and M.~N. Kenari, ``{Type Based Sign
  Modulation and its Application for ISI mitigation in Molecular
  Communication},'' \emph{IEEE Trans. Commun.}, 2017.

\bibitem{StochasticReaction_Kampen}
N.~V. Kampen, \emph{Stochastic Processes in Physics and Chemistry}, third
  edition~ed., ser. North-Holland Personal Library.\hskip 1em plus 0.5em minus
  0.4em\relax Amsterdam: Elsevier, 2007.

\bibitem{TableIntegSerie}
I.~S. Gradshteyn and I.~M. Ryzhik, \emph{Table of Integrals, Series, and
  Products}.\hskip 1em plus 0.5em minus 0.4em\relax 7th ed. Academic, 2007.

\bibitem{PoissonGardiner}
C.~Gardiner and S.~Chaturvedi, ``{The Poisson Representation. I. A New
  Technique for Chemical Master Equations},'' \emph{Journal of Statistical
  Physics}, vol.~17, no.~6, pp. 429--468, 1977.

\bibitem{CoxNatureCommun}
D.~Schnoerr, R.~Grima, and G.~Sanguinetti, ``{Cox Process Representation and
  Inference for Stochastic Reaction-Diffusion Processes},'' \emph{Nature
  Commun.}, vol.~7, 2016.

\bibitem{ReactionDiffSim}
S.~S. Andrews and D.~Bray, ``{Stochastic Simulation of Chemical Reactions with
  Spatial Resolution and Single Molecule Detail},'' \emph{Physical biology},
  vol.~1, no.~3, p. 137, 2004.

\bibitem{Adam_AcCoRD}
A.~Noel, K.~C. Cheung, R.~Schober, D.~Makrakis, and A.~Hafid, ``{Simulating
  with AcCoRD: Actor-based Communication via Reaction-Diffusion},'' \emph{Nano
  Commun. Netw.}, vol.~11, pp. 44 -- 75, 2017.

\end{thebibliography}


\begin{thebibliography}{10}
\providecommand{\url}[1]{#1}
\csname url@samestyle\endcsname
\providecommand{\newblock}{\relax}
\providecommand{\bibinfo}[2]{#2}
\providecommand{\BIBentrySTDinterwordspacing}{\spaceskip=0pt\relax}
\providecommand{\BIBentryALTinterwordstretchfactor}{4}
\providecommand{\BIBentryALTinterwordspacing}{\spaceskip=\fontdimen2\font plus
\BIBentryALTinterwordstretchfactor\fontdimen3\font minus
  \fontdimen4\font\relax}
\providecommand{\BIBforeignlanguage}[2]{{%
\expandafter\ifx\csname l@#1\endcsname\relax
\typeout{** WARNING: IEEEtran.bst: No hyphenation pattern has been}%
\typeout{** loaded for the language `#1'. Using the pattern for}%
\typeout{** the default language instead.}%
\else
\language=\csname l@#1\endcsname
\fi
#2}}
\providecommand{\BIBdecl}{\relax}
\BIBdecl

\bibitem{Nariman_Survey}
N.~Farsad, H.~Yilmaz, A.~Eckford, C.~Chae, and W.~Guo, ``{A Comprehensive
  Survey of Recent Advancements in Molecular Communication},'' \emph{IEEE
  Commun. Surveys Tutorials}, vol.~18, no.~3, pp. 1887--1919, third quarter
  2016.

\bibitem{CellBio}
B.~Alberts, D.~Bray, K.~Hopkin, A.~Johnson, J.~Lewis, M.~Raff, K.~Roberts, and
  P.~Walter, \emph{{Essential Cell Biology}}.\hskip 1em plus 0.5em minus
  0.4em\relax New York, NY: Garland Science, 4th ed., 2014.

\bibitem{Adam_Enzyme}
A.~Noel, K.~Cheung, and R.~Schober, ``{Improving Receiver Performance of
  Diffusive Molecular Communication with Enzymes},'' \emph{IEEE Trans.
  NanoBiosci.}, vol.~13, no.~1, pp. 31--43, Mar. 2014.

\bibitem{Nariman_Acid}
N.~Farsad and A.~Goldsmith, ``{A Novel Molecular Communication System Using
  Acids, Bases and Hydrogen Ions},'' in \emph{Proc. IEEE SPAWC}, Jul. 2016, pp.
  1--6.

\bibitem{Nariman_AcidBasePlatform}
\BIBentryALTinterwordspacing
N.~Farsad, D.~Pan, and A.~Goldsmith, ``{A Novel Experimental Platform for
  In-Vessel Multi-Chemical Molecular Communications},'' \emph{Accepted for
  presentation in IEEE Globecom}, 2017. [Online]. Available:
  \url{https://arxiv.org/abs/1704.04810}
\BIBentrySTDinterwordspacing

\bibitem{NonlinearPDE_Debnath}
L.~Debnath, \emph{{Nonlinear Partial Differential Equations for Scientists and
  Engineers}}.\hskip 1em plus 0.5em minus 0.4em\relax Springer Science \&
  Business Media, 2011.

\bibitem{PDE_numerical}
J.~Lang, ``{Numerical Solution of Reaction-Diffusion Equations},'' in
  \emph{Scientific Computing in Chemical Engineering}.\hskip 1em plus 0.5em
  minus 0.4em\relax Springer, 1996, pp. 136--141.

\bibitem{ReactionDiffSim}
S.~S. Andrews and D.~Bray, ``{Stochastic Simulation of Chemical Reactions with
  Spatial Resolution and Single Molecule Detail},'' \emph{Physical biology},
  vol.~1, no.~3, p. 137, 2004.

\bibitem{Chou_MasterEq}
C.~T. Chou, ``{Extended Master Equation Models for Molecular Communication
  Networks},'' \emph{IEEE Trans. NanoBiosci.}, vol.~12, no.~2, pp. 79--92, Jun.
  2013.

\bibitem{Adam_AcCoRD}
A.~Noel, K.~C. Cheung, R.~Schober, D.~Makrakis, and A.~Hafid, ``{Simulating
  with AcCoRD: Actor-based Communication via Reaction-Diffusion},'' \emph{Nano
  Commun. Netw.}, vol.~11, pp. 44 -- 75, 2017.

\bibitem{Reza_Reaction}
R.~Mosayebi, A.~Gohari, M.~Mirmohseni, and M.~N. Kenari, ``{Type Based Sign
  Modulation and its Application for ISI Mitigation in Molecular
  Communication},'' \emph{IEEE Trans. Commun.}, 2017.

\bibitem{ICC2017_MC_IEEE}
V.~Jamali, A.~Ahmadzadeh, and R.~Schober, ``{Symbol Synchronization for
  Diffusive Molecular Communications},'' in \emph{Proc. IEEE ICC}, May 2017.

\bibitem{PoissonGardiner}
C.~Gardiner and S.~Chaturvedi, ``{The Poisson Representation. I. A New
  Technique for Chemical Master Equations},'' \emph{Journal of Statistical
  Physics}, vol.~17, no.~6, pp. 429--468, 1977.

\bibitem{CoxNatureCommun}
D.~Schnoerr, R.~Grima, and G.~Sanguinetti, ``{Cox Process Representation and
  Inference for Stochastic Reaction-Diffusion Processes},'' \emph{Nature
  Commun.}, vol.~7, 2016.

\bibitem{TCOM_MC_CSI}
V.~Jamali, A.~Ahmadzadeh, C.~Jardin, C.~Sticht, and R.~Schober, ``{Channel
  Estimation for Diffusive Molecular Communications},'' \emph{IEEE Trans.
  Commun.}, vol.~64, no.~10, pp. 4238--4252, Oct. 2016.

\bibitem{Gau_AB}
S.~Wang, W.~Guo, and M.~D. McDonnell, ``{Transmit Pulse Shaping for Molecular
  Communications},'' in \emph{IEEE Infocom Workshops}, Apr. 2014, pp. 209--210.

\bibitem{NanoCOM16}
V.~Jamali, N.~Farsad, R.~Schober, and A.~Goldsmith, ``{Non-Coherent
  Multiple-Symbol Detection for Diffusive Molecular Communications},'' in
  \emph{Proc. ACM NanoCom}, Sept. 2016.

\end{thebibliography}

\end{document}